\documentclass[aip,jcp,reprint,showkeys]{revtex4-1}

\usepackage{graphicx,bm,xcolor,amsmath,amssymb,amsfonts,float,physics}

\usepackage{algorithmicx}
\usepackage{algcompatible}
\usepackage{algpseudocode}
\newcommand{\LeftComment}[1]{\State {\scriptsize /* \textit{#1} */}}

\usepackage[utf8]{inputenc}
\usepackage[T1]{fontenc}
\usepackage{txfonts}

\usepackage[
    colorlinks=true,
    citecolor=blue,
    linkcolor=blue,
    filecolor=blue,      
    urlcolor=blue,	
    breaklinks=true
	]{hyperref}
\newcommand{\stwo}{\hat{S}^2}
\newcommand{\szop}{\hat{S}_z}
\newcommand{\tu}{\mathtt{u}}

\newcommand{\md}{\mathtt{d}}
\newcommand{\mpp}{\mathtt{p}}
\newcommand{\mpv}{\mathbf{p}}
\newcommand{\up}{\uparrow}
\newcommand{\dn}{\downarrow}
\newcommand{\Nint}{{N_\text{int}}}
\newcommand{\Norb}{{N_\text{orb}}}
\newcommand{\one}{{\texttt{1}}}
\newcommand{\zero}{{\texttt{0}}}

\newcommand{\cipsi}{CIPSI}
\newcommand{\csf}{CSF}
\newcommand{\sci}{SCI}
\newcommand{\cpu}{CPU}
\newcommand{\ept}{E_\text{PT2}}

\makeatletter
\newcommand{\superimpose}[2]{%
  {\ooalign{$#1\@firstoftwo#2$\cr\hfil$#1\@secondoftwo#2$\hfil\cr}}}
\makeatother

\newcommand{\orbup}{\mathpalette\superimpose{{-}{\uparrow}}}
\newcommand{\orbdn}{\mathpalette\superimpose{{-}{\downarrow}}}
\newcommand{\orb}{\mathpalette\superimpose{{-}{}}}
\newcommand{\orbocc}{\mathpalette\superimpose{{-}{|}}}
\newcommand{\orboccc}{\mathpalette\superimpose{{-}{||}}}

\begin{document}

\title{Spin-adapted selected configuration interaction in a determinant basis}

\author{Vijay Gopal Chilkuri}
\affiliation{Laboratoire de Chimie et Physique Quantiques, Universit\'e de Toulouse, CNRS, UPS, France}
\author{Thomas Applencourt}
\affiliation{Argonne Leadership Computing Facility, Argonne National Laboratory, Argonne, IL 60439 USA}
\author{Kevin Gasperich}
\affiliation{Argonne Leadership Computing Facility, Argonne National Laboratory, Argonne, IL 60439 USA}
\affiliation{Department of Chemistry, University of Pittsburgh, Pittsburgh, Pennsylvania 15260 USA}
\author{Pierre-Fran\c{c}ois Loos}
\affiliation{Laboratoire de Chimie et Physique Quantiques, Universit\'e de Toulouse, CNRS, UPS, France}
\author{Anthony Scemama}
\affiliation{Laboratoire de Chimie et Physique Quantiques, Universit\'e de Toulouse, CNRS, UPS, France}
\email{scemama@irsamc.ups-tlse.fr}


\begin{abstract}
  Selected configuration interaction ({\sci}) methods, when complemented with a
  second-order perturbative correction, provide near full configuration
  interaction (FCI) quality energies with only a small fraction of the Slater
  determinants of the FCI space. However, a selection criterion based on
  determinants alone does not ensure a spin-pure wave function.  In other words,
  such {\sci} wave functions are not eigenfunctions of the $\stwo$ operator. In
  many situations (bond breaking, magnetic system, excited state, etc), having a 
  spin-adapted wave function is essential for a quantitatively correct description 
  of the system. Here, we propose an efficient algorithm which,
  given an arbitrary determinant space, generates all the missing Slater
  determinants allowing one to obtain spin-adapted wave functions while avoiding
  manipulations involving configuration state functions. For example, generating all the
  possible determinants with 6 spin-up and 6 spin-down electrons in 12 open
  shells takes 21 {\cpu} cycles per generated Slater determinant. The
  selection is still done with individual determinants, and one can
  take advantage of the basis of configuration state functions in the
  diagonalization of the Hamiltonian to reduce significantly the memory
  footprint.
\end{abstract}

\maketitle

\section{Introduction}

In recent years, selected configuration interaction ({\sci}) methods \cite{whitten_1969,bender_1969,huron_1973} have become more and more popular,\cite{greer_1998,hanrath_1997,stampfuss_2005,bytautas_2009,booth_2009,giner_2013,buenker_2014,holmes_2016,liu_2014,liu_2016,lei_2017,zhang_2020,zimmerman_2017,ohtsuka_2017,coe_2018,scemama_2018a,evangelista_2014,flad_2010,eriksen_2020,garniron_2018,loos_2020,scemama_2020,benali_2020,schriber_2016,li_2018,yao_2020,li_2020,williams_2020}
especially for the accurate calculation of electronic excitation
energies.\cite{coe_2013,schriber_2017,holmes_2017,loos_2018,scemama_2018,dash_2018,chien_2018,loos_2019,loos_2020b,loos_2020c,loos_2020d,veril_2021,dash_2019,giner_2019,scemama_2019,loos_2019,blunt_2015}
\emph{Determinant-based {\sci}}
refers to configuration interaction in a truncated space of
determinants.  For instance, a SCI with singles and
doubles (SCISD) refers to the diagonalization of the CISD
Hamiltonian with only a subset of chosen (or selected) determinants belonging to the CISD space.
There exists many variants of {\sci} methods differing in two major
aspects. The first one is the nature of the target space: the most
common spaces are the multi-reference CI (MRCI)
space,\cite{hanrath_1997,neese_2003,buenker_2014} the (frozen-core) full CI (FCI)
space\cite{greer_1998,booth_2009,giner_2013,holmes_2017,evangelista_2014}
and the complete active space (CAS).\cite{smith_2017} The second
aspect in which {\sci} methods differ are in the rules used to select
the determinants, thus affecting the convergence with respect to the
number of determinants and the computational cost. The various permutations
of such rules results in a plethora of {\sci} methods.

Discussing the different kinds of selection rules is beyond the scope
of the present article. The reader who is not acquainted with {\sci}
methods only needs to be aware of a few key aspects:
(i) the selection criterion is chosen to include the most
energetically relevant determinants in the variational space;
(ii)  {\sci} methods produce wave functions that are potentially expanded in an \emph{arbitrary} set of determinants;
(iii)  it is of common practice to compute the Epstein-Nesbet second-order
  perturbative correction ($\ept$) to the variational energy, in order to
  estimate the lowest eigenvalues of the CI Hamiltonian matrix
  defined by the method (CAS, MRCI, FCI, \dots);
(iv) as the number of determinants grows, $\ept \rightarrow 0$ and the variational {\sci} energy converges monotonically to the exact energy of the CI Hamiltonian.

A balanced description of excited states, magnetic systems, and
bond breakings require the wave functions to be spin-adapted,
i.e., eigenfunctions of the $\stwo$ operator.
The Slater determinant many-particle representation is, by construction, only
strictly an eigenfunction of the $\szop$ operator and therefore does not
ensure a spin-pure wave function.
The usual way to enforce the wave function to be an eigenfunction
of $\stwo$ is to work in a basis where each element of the basis is an
eigenfunction of $\stwo$ with the desired eigenvalue. These basis functions
are built as linear combinations of Slater determinants, and are known
as \emph{configuration state functions} ({\csf}).

A natural option would be to express {\sci} in terms of {\csf}s.
However, due to the complexity in the calculation of matrix elements
in the {\csf} basis, many {\sci} implementations still rely on
determinants. Opting for the {\csf}
representation would require a major effort for re-writing the software,
such as the important work that was done in the NECI FCIQMC code which
now uses the graphical unitary group
approach\cite{paldus_2021a,paldus_2021b,paldus_2021c,dobrautz_2019,limanni_2020}
and the ORCA program which uses
the angular-momentum coupling based approach\cite{pipano1968convergence,shavitt1977graph,paldus1976theoretical,paldus1975pattern,paldus1976unitary,vijay1_2020,vijay2_2020}.
In the present paper we follow a different route and present simple recipes to
ensure that the selected wave functions are spin-adapted without
requiring too many modifications in a determinant-based code.

\section{Many-particle basis representations}

A \emph{configuration} is a vector of molecular orbital occupation
numbers. For example, the configuration $(2,1,1,1,1)$ can be written
as a linear combination of six determinants
\begin{equation}
\left( {\scriptsize \begin{array}{c} \orbocc \\ \orbocc \\ \orbocc \\ \orbocc \\ \orboccc \end{array} } \right)
 =  
 a \left( {\scriptsize \begin{array}{c} \orb \orbdn  \\ \orb \orbdn  \\ \orbup \orb   \\ \orbup \orb \\ \orbup \orbdn \end{array} } \right) 
 + b \left( {\scriptsize \begin{array}{c} \orbup \orb   \\ \orb \orbdn \\ \orb \orbdn  \\ \orbup \orb  \\ \orbup \orbdn \end{array} } \right) 
 + c \left( {\scriptsize \begin{array}{c} \orbup \orb   \\ \orb \orbdn \\ \orbup \orb  \\ \orb \orbdn  \\ \orbup \orbdn \end{array} } \right)
 + d \left( {\scriptsize \begin{array}{c} \orbup \orb   \\ \orbup \orb \\ \orb \orbdn  \\ \orb \orbdn  \\ \orbup \orbdn \end{array} } \right) 
 + e \left( {\scriptsize \begin{array}{c} \orb \orbdn   \\ \orbup \orb \\ \orbup \orb  \\ \orb \orbdn  \\ \orbup \orbdn \end{array} } \right) 
 + f \left( {\scriptsize \begin{array}{c} \orb \orbdn   \\ \orbup \orb \\ \orb \orbdn  \\ \orbup \orb  \\ \orbup \orbdn \end{array} } \right) 
  \label{eq:det}
\end{equation}
or of two {\csf s} with coefficients $A$ and $B$
\begin{widetext}
\begin{equation}
\left( {\scriptsize \begin{array}{c} \orbocc \\ \orbocc \\ \orbocc \\ \orbocc \\ \orboccc \end{array} } \right)
 = A \times \frac{1}{2} \left[
   \left( {\scriptsize \begin{array}{c} \orbup \orb   \\ \orb \orbdn \\ \orb \orbdn  \\ \orbup \orb  \\ \orbup \orbdn \end{array} } \right)
  + \left( {\scriptsize \begin{array}{c} \orb \orbdn   \\ \orbup \orb \\ \orbup \orb  \\ \orb \orbdn  \\ \orbup \orbdn \end{array} } \right)
  - \left( {\scriptsize \begin{array}{c} \orbup \orb   \\ \orb \orbdn \\ \orbup \orb  \\ \orb \orbdn  \\ \orbup \orbdn \end{array} } \right)
  - \left( {\scriptsize \begin{array}{c} \orb \orbdn   \\ \orbup \orb \\ \orb \orbdn  \\ \orbup \orb  \\ \orbup \orbdn \end{array} } \right) 
\right]
 + B \times \frac{\sqrt{3}}{6} \left[ 
 -2 \left( {\scriptsize \begin{array}{c} \orb \orbdn  \\ \orb \orbdn  \\ \orbup \orb   \\ \orbup \orb \\ \orbup \orbdn \end{array} } \right) 
  + \left( {\scriptsize \begin{array}{c} \orbup \orb   \\ \orb \orbdn \\ \orb \orbdn  \\ \orbup \orb  \\ \orbup \orbdn \end{array} } \right) 
  + \left( {\scriptsize \begin{array}{c} \orbup \orb   \\ \orb \orbdn \\ \orbup \orb  \\ \orb \orbdn  \\ \orbup \orbdn \end{array} } \right)
 -2 \left( {\scriptsize \begin{array}{c} \orbup \orb   \\ \orbup \orb \\ \orb \orbdn  \\ \orb \orbdn  \\ \orbup \orbdn \end{array} } \right) 
  + \left( {\scriptsize \begin{array}{c} \orb \orbdn   \\ \orbup \orb \\ \orbup \orb  \\ \orb \orbdn  \\ \orbup \orbdn \end{array} } \right) 
  + \left( {\scriptsize \begin{array}{c} \orb \orbdn   \\ \orbup \orb \\ \orb \orbdn  \\ \orbup \orb  \\ \orbup \orbdn \end{array} } \right)
\right]
  \label{eq:csf}
\end{equation}
\end{widetext}
By definition, all the determinants belonging to the same {\csf} are
associated with the same configuration, and the determinants associated
with a given configuration may be involved in multiple {\csf}s.
Expressing Eq.~\eqref{eq:det} in terms of
{\csf}s is an overdetermined problem: six parameters $(a,b,c,d,e,f)$ for determinants
\textit{vs} two parameters $(A,B)$ for the {\csf}s, so it has no unique solution in the
general case. Only eigenfunctions of the $\stwo$ operator possess the
necessary constraints to enable the exact transformation.

A few years ago, Bytautas and Ruedenberg proposed a simple scheme to
truncate large determinant-based wave functions while maintaining the
spin purity.\cite{bytautas_2007}
The squared coefficients of
the determinants within the same configuration are summed
together to produce the so-called \emph{space-product weights}, which
are then used to truncate the wave function.
As the truncation occurs by removing configurations,
one can understand from Eqs.~\eqref{eq:det} and~\eqref{eq:csf} that
the removal of all the determinants associated with a configuration
is equivalent to a removal of all the CSFs associated with the same
configuration, hence keeping the spin purity of the wave function.

Following this idea, imposing spin adaptation in determinant-based
{\sci} methods can be done by
(i) identifying all the configurations of the determinants composing
  the variational space,
(ii) generating all the determinants of the required multiplicity corresponding to these
  configurations, and
(iii) diagonalizing the Hamiltonian in this spin-complete determinant
  space.
Since the Hamiltonian commutes with the $\stwo$ operator, the obtained
eigenfunction is automatically spin-adapted. An efficient algorithm to carry out
this procedure is presented in this paper. Because the obtained 
wave functions are spin-adapted, they can be exactly expressed in terms of
{\csf}s.\cite{grabenstetter1976generation,pauncz2012spin,olsen_2014,fales_2020} Then, we take advantage of the
reduction of the number of parameters to reduce the memory requirement
of the Davidson diagonalization, which is the main bottleneck in today's
{\sci} algorithms.
All the presented algorithms are implemented in the open-source \emph{Quantum
Package} software.\cite{garniron_2019}

\section{Algorithm}

The wave function of a given electronic state is expressed as
$ \ket{\Psi} = \sum_I c_I \ket{D_I} $,
where each Slater determinant $D_I$ is represented as a Waller-Hartree
double determinant,\cite{pauncz_1989}
$ D_I = d_i^\up \, d_j^\dn $,
i.e., the product of a determinant of spin-up ($\up$) orbitals $d_i^\up$ and a determinant
of spin-down ($\dn$) orbitals $d_j^\dn$.  Such a representation can be encoded as a
pair of bit strings $(\md_i,\md_j)$, where each bit string is of length
$\Norb$, the number of molecular orbitals.  The spin-up and spin-down orbitals
originate from a restricted Hartree-Fock or a CAS self-consistent field (CASSCF) calculation, so that the 
spatial part of these orbitals are common for both spin manifolds.  Within
a bit string, each bit corresponds to a spin-orbital; the bit is set
to $\one$ if the orbital is occupied, and it is set to $\zero$ if the
orbital is empty.  In low-level languages such as Fortran or C, a bit
string may be stored as an array of $\Nint$ 64-bit integers, where
\begin{equation}
  \Nint = \left \lfloor \frac{\Norb-1}{64} \right \rfloor + 1
\end{equation}

This representation allows for efficient determinant comparisons using
bit-wise operation capabilities of modern
processors\cite{scemama_2013} and will be convenient in the following.

All the {\cpu} cycle measurements were performed on an Intel(R)
Xeon(R) Gold 6140 \cpu @2.30GHz with the GNU Fortran compiler 7.3.0,
by reading the time stamp counter of the {\cpu} with the
\texttt{rdtsc} instruction.

\subsection{Identification of the configurations}

The configuration $\mpv_I$ associated with determinant $D_I$
is a vector of integers defined as
\begin{equation}
  [\mpv_I]_k =
  \begin{cases}
    0 & \text{when the $k$-th orbital is unoccupied} \\
    1 & \text{when the $k$-th orbital is singly occupied} \\
    2 & \text{when the $k$-th orbital is doubly occupied}
  \end{cases}
\end{equation}
If $\mpv_I$ is encoded as a pair of bit strings
$(\mpp_I^{(1)}, \mpp_I^{(2)})$, where $\mpp_I^{(1)}$ and
$\mpp_I^{(2)}$ encode respectively the singly and doubly occupied
orbitals, the configuration can be computed as
\begin{equation}
  \label{eq:sop}
    \mpp_I^{(1)}  = \md_i \oplus \md_j \;\;\text{ and }\;\;
    \mpp_I^{(2)}  = \md_i \wedge \md_j
\end{equation}
where $\oplus$ and $\wedge$ denote respectively the \texttt{xor} and 
the \texttt{and} binary operators.

Transforming all the selected determinants into a list
of unique configurations can be done in linear time if a hash value is
associated with each configuration.\cite{bitton_1983} Hence, the time
for this transformation is negligible.

\subsection{Generating all the determinants associated with a configuration}

Given a configuration, one must generate all the possible
determinants by considering either a spin-up or a spin-down electron 
in the singly occupied molecular orbitals,
keeping the numbers of spin-up and spin-down electrons fixed.  One can
notice that, by doing so, all the generated determinants only differ by these
singly occupied orbitals, so from now on we can consider a more
compact representation: a bit string of $n_\up + n_\dn$ bits, where
$n_\up$ and $n_\dn$ denote the numbers of spin-up and spin-down unpaired
electrons. The bit is set to $\one$ when the orbital is occupied by a
spin-up electron, and $\zero$ when it is occupied by a spin-down electron.
The indices of the singly occupied orbitals are kept in a look-up
table $\mathbf{m}$ for later use.

\begin{figure}[ht]
  \begin{algorithmic}
\Function{compute\_permutations}{$n,m$} \LeftComment{$\mathtt{n}$:
  input, number of bits set to $\one$} \LeftComment{$\mathtt{m}$:
  input, number of bits set to $\zero$} \LeftComment{$\mathtt{v}$:
  output, an array of permutations} \LeftComment{$\mathtt{u}$,
  $\mathtt{t}$, $\mathtt{t'}$, $\mathtt{t''}$ and $\mathtt{v}$ are
  encoded in at least $\mathtt{n+m+1}$ bits} \State
$\mathtt{k \gets 0}$ \State $\mathtt{u \gets (1 \ll n) - 1}$
\While {$\mathtt{u < \left(1 \ll (n+m) \right) }$} \State
$\mathtt{v[k] \gets u}$ \State $\mathtt{k \gets k+1}$ \State
$\mathtt{t \gets u \vee (u-1)}$ \State $\mathtt{t' \gets t + 1}$\
\State
$\mathtt{t'' \gets \left((\neg t \wedge t')-1 \right) \gg
  (ctz(u)+1)}$ \State $\mathtt{u \gets t' \vee t''}$ \EndWhile
\State \Return $\mathtt{v}$ \EndFunction
  \end{algorithmic}
  \begin{tabular}{ccccc}
\hline
Iteration & \texttt{t} & \texttt{t}' & \texttt{t}'' & \texttt{u} \\
\hline
0 &               &               &               & \texttt{0011} \\
1 & \texttt{0011} & \texttt{0100} & \texttt{0001} & \texttt{0101} \\
2 & \texttt{0101} & \texttt{0110} & \texttt{0000} & \texttt{0110} \\
3 & \texttt{0111} & \texttt{1000} & \texttt{0001} & \texttt{1001} \\
4 & \texttt{1001} & \texttt{1010} & \texttt{0000} & \texttt{1010} \\
5 & \texttt{1011} & \texttt{1100} & \texttt{0000} & \texttt{1100} \\
\hline
  \end{tabular}

  \caption{ Anderson's algorithm. All the configurations of
    $\mathtt{n}$ bits set to $\one$ are generated in an integer of $\mathtt{n+m}$
    bits in lexicographic order.  $\mathtt{ctz(i)}$ counts the number
    of trailing zeros, $\mathtt{i \ll n}$ shifts $\mathtt{i}$ by
    $\mathtt{n}$ bits to the left, $\mathtt{i \gg n}$ shifts
    $\mathtt{i}$ by $\mathtt{n}$ bits to the right, $\mathtt{\wedge}$
    is the bit-wise \texttt{and} operator, and $\mathtt{\vee}$ is the
    bit-wise \texttt{or} operator.}
  \label{fig:algo}
\end{figure}

To generate all the determinants keeping the numbers of spin-up and
spin-down electrons constant, we need to build all the possible bit
strings with $n_\up$ bits set to $\one$ and $n_\dn$ bits set to
$\zero$.  This compact representation allows us to use Anderson's
algorithm (see Fig.~\ref{fig:algo}),\cite{NextBit} which generates all the configurations of
$n_\up$ bits set to $\one$ in a bit string of length $n_\up+n_\dn$ in
lexicographical order.
To illustrate how this algorithm proceeds, we also show the
step-by-step transformations of the variables with $n_\up=2$ and
$n_\dn=2$ from which the sequence \texttt{(0011, 0101, 0110, 1001,
1010, 1100)} is produced.

\begin{figure}[ht]
  \begin{center}
  \includegraphics[width=\linewidth]{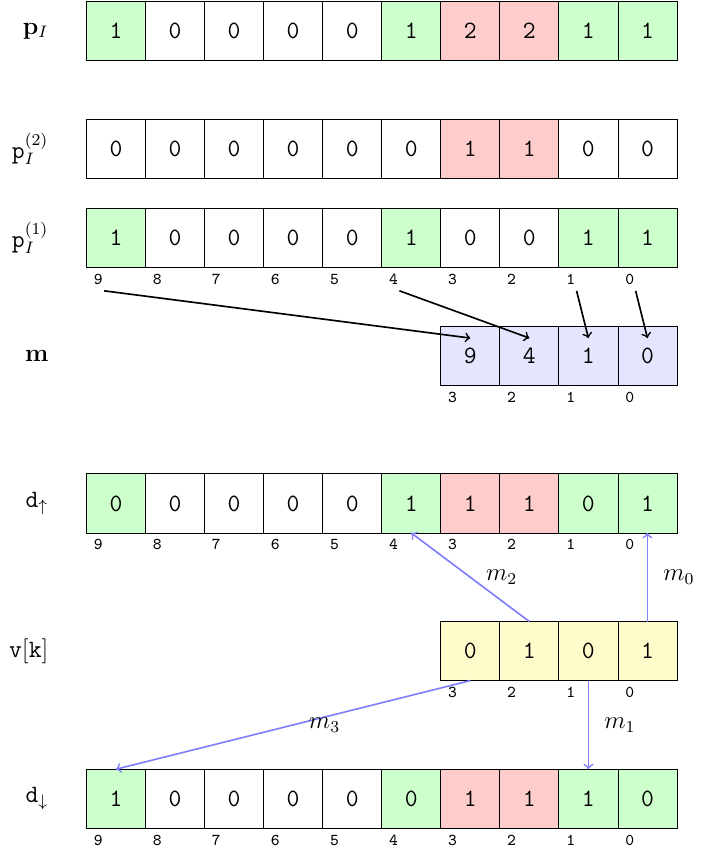}
  \end{center}
  \caption{The configuration $\mpv_I$ is encoded as in
    Eq.~\eqref{eq:sop}. Singly and doubly occupied orbitals are
    represented respectively in green and red.  The list of indices
    $\mathbf{m}$ of the singly occupied orbitals is built (in blue),
    and this mapping is re-used to build the determinants from
    permutations (yellow) generated by Anderson's algorithm.
    Bit strings and arrays are represented from right to left to be
    consistent with the binary notation.}
  \label{fig:mapping}
\end{figure}

Figure \ref{fig:mapping} gives a pictorial description of the data
structures used to generate a determinant.  To build a generated
determinant $(\md_\up,\md_\dn)$ from a permutation $\tu$, one must
first fill the doubly occupied orbitals by setting both $\md_\up$ and
  $\md_\dn$ equal to $\mpp_I^{(2)}$.
Then, one must iterate over the bits of $\tu$. If the $k$-th bit is set to
  $\one$, set the $m_k$-th orbital of $d_\up$ to $\one$, otherwise set
  the $m_k$-th orbital of $d_\dn$ to $\one$.

\subsection{Further optimizations}

As a first optimization, instead of creating each determinant from the
permutation as shown in Fig.~\ref{fig:mapping}, all the determinants
can be generated iteratively by considering only the orbitals that
have changed from the previously generated determinant. This avoids
always setting all the $n_\up+n_\dn$ bits in the bit strings.  The
integer obtained by $\mathtt{v[k-1]} \oplus \mathtt{v[k]}$ has bits
set to $\one$ at the positions where the bits differ between
$\mathtt{v[k-1]}$ and $\mathtt{v[k]}$. The positions of these bits can
be found in a few cycles by first 
counting the number of trailing zeros (this gives the position
  of the least significant $\one$),
then by Setting the least significant $\one$ to $\zero$ using
  $\mathtt{v[k] \gets v[k] \wedge (v[k]-1)}$
and iterating until $\mathtt{v[k]} = 0$.

A second optimization is to consider time-reversal symmetry
(i.e., exchanging all spin-up and spin-down electrons in an even electron systems).
When $n_\up = n_\dn$, one can remark that
${\mathtt{v[n_{det}}-1-\mathtt{k]} = \neg \mathtt{v[k]}}$, where
$\mathtt{n_{det}}$ is the number of generated determinants:
$ \mathtt{n_{det}} = \frac{(n_\up +n_\dn)!}{n_\up! n_\dn!}$.
Hence, it is sufficient to iterate over the first half of the permutations
of Anderson's algorithm, and generate pairs of determinants per iteration.

\subsection{Reduction of the memory requirements}

In the latest version of \emph{Quantum Package}, the spin-pure
eigenstates were obtained by finding the lowest eigenstate of a linear combination of the
Hamiltonian and the $\stwo$ matrices.\cite{garniron_2019,fales_2017}
At iteration $n$, the Davidson algorithm requires the computation
of the matrix $\mathbf{W = H\, U}$, where $\mathbf{U}$ and
$\mathbf{W}$ are $N_{\text{det}} \times N_{\text{states}}$ matrices, where
$N_\text{det}$ is the number of determinants, and $N_{\text{states}}$ is
greater than the number of states of interest and adjusted to
reduce the number of iterations for the convergence of the algorithm.

In terms of storage, the $\mathbf{W}$ and $\mathbf{U}$ matrices of all $n$ 
iterations need to be stored. As the storage increases with
iterations, it is common practice to define a maximum iteration
$n_{\text{max}}$ where all the $\mathbf{U}$ matrices are compressed
into a single $N_{\text{det}} \times N_{\text{states}}$ improved
$\mathbf{U}$ matrix, and the algorithm restarts.
If one wants to monitor the expectation value $\expval{\stwo}$,
one needs also to compute $\mathbf{Y} = \mathbf{S^2\,U}$, and store
the $\mathbf{Y}$ matrices of all iterations.
As the computation of $\stwo$ is made only
for monitoring purposes, the $\mathbf{Y}$ matrices can be stored in
single precision to limit the increase in the memory requirements.
Hence, in the determinant basis, the required space for the
diagonalization is
$2.5 \times N_{\text{det}} \times N_{\text{states}} \times n_{\text{max}}$.

Since the selected determinant space contains all the determinants of
each configuration, we can make an exact transformation from the
determinant basis to the CSF basis,\cite{olsen_2014,fales_2020}
thus rendering the many-particle basis representation more compact. Hence, we
now store the $\mathbf{U}$ and $\mathbf{W}$ matrices in the CSF basis,
while the computation $\mathbf{W = H\, U}$ is still performed in the
determinant basis for simplicity. As the wave function is guaranteed
to be an eigenstate of $\stwo$, it is no longer necessary to compute
and store $\mathbf{S^2}$, so the storage requirements are reduced to
$2 \times N_{\text{det}} \times N_{\text{states}} + 
2 \times N_{\text{CSF}} \times N_{\text{states}} \times n_{\text{max}}$, 
where $N_{\text{CSF}} \ll N_{\text{det}}$ is the number of CSFs (see below).

\section{Numerical tests}

In this section, the \textit{configuration interaction using a perturbative 
selection made iteratively} ({\cipsi}) algorithm \cite{huron_1973,garniron_2019}
is employed to select determinants of the
external space: they are selected by the magnitude of their
contribution to the second-order perturbative correction to the
energy.
The spin-adaptation step is introduced between the selection step and
the diagonalization.
We would like to emphasize that we use {\cipsi} because it is the
method implemented in \emph{Quantum Package}, but any CI or {\sci}
could be have been considered.

\subsection{Avoided crossing of LiF}

The avoided crossing between the ionic and neutral $^1\Sigma^+$ states
of LiF is a common benchmark for correlated methods, as the location
of the crossing is highly sensitive to the amount of
correlation. \cite{casanova_2012,legeza_2003,garniron_2017a}
At large distances, the lowest triplet state is very close in energy to
the singlet states. If the wave function is not spin-adapted, the
triplet state will mix with the singlets during the selection, and the
convergence of the {\cipsi} calculation to the correct states is not
guaranteed.

\begin{figure}[ht]
  \begin{center}
  \includegraphics[width=\columnwidth]{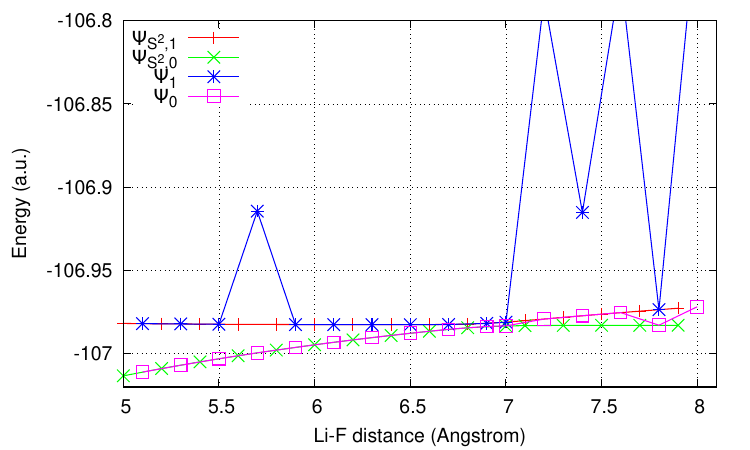}
  \end{center}
  \caption{Avoided crossing of LiF, with ($\Psi_{S^2,0}$,
$\Psi_{S^2,1}$) and without ($\Psi_0$, $\Psi_1$) imposing spin
symmetry. The energy (in hartree) of the two lowest singlet states of
LiF is represented as a function of the bond length (in \AA).}
  \label{fig:lif}
\end{figure}

We report in Fig.~\ref{fig:lif} the potential energy curve of the two
lowest singlet states of LiF computed with and without imposing spin
adaptation. For all the distances, the {\cipsi} calculations were run
\emph{blindly} (with no user interaction), starting with the
CASSCF(2,2)/aug-cc-pVDZ wave functions of both states (four
determinants). Only the lowest molecular orbital was frozen,
corresponding to the $1s$ orbital of the fluorine atom. The
calculations were stopped when the second-order perturbative
correction was below $0.1$~m$E_h$ or when the number of determinants
reached 4 million.

Figure \ref{fig:lif} shows that for large distances, without spin
adaptation, there are multiple erratic points for which the two
obtained states are not the desired ones. This curve also shows that
all the points obtained with spin-adaptation converged to the correct
states, giving a smooth potential energy curve.

\subsection{Dissociation of N$_2$}

Selected CI methods provide not only the energies of the states of
interest, but also the corresponding wave functions which can be used
for post-processing. For instance, wave functions computed with CIPSI
have shown to be excellent choices of trial wave functions for quantum
Monte Carlo calculations.\cite{caffarel_2016,scemama_2018a,scemama_2018,scemama_2019,scemama_2020,benali_2020} 
When a wave function is used for further calculations, the spin-adapted characteristic is
particularly important because it can enforce a continuous
behavior of wave function along a dissociation curve, especially when
different spin states become quasi-degenerate.

\begin{figure*}[ht]
  \begin{center}
  \includegraphics[width=0.48\textwidth]{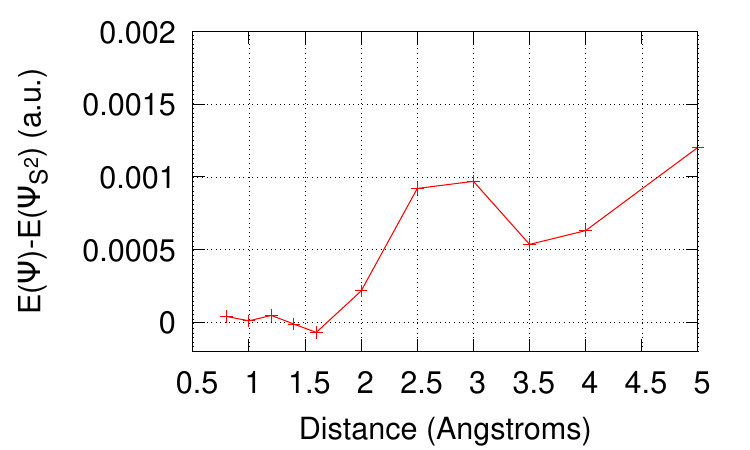}
  \includegraphics[width=0.48\textwidth]{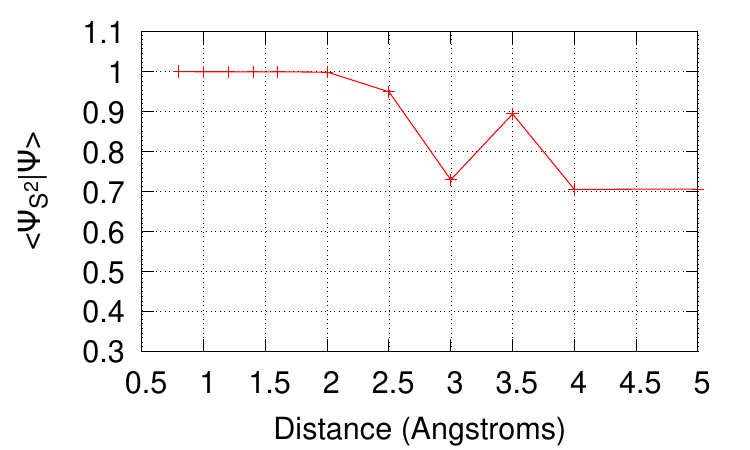}
  \end{center}
  \caption{Difference of extrapolated FCI energy obtained with
    ($\Psi_{S^2}$) and without ($\Psi$) spin-adaptation along the
dissociation path of N$_2$ (aug-cc-pVDZ). The overlaps of the wave
functions obtained with the two schemes are reported on the right.}
  \label{fig:n2}
\end{figure*}

To illustrate the importance of this feature, we compute the
dissociation curve of the singlet ground state of the N$_2$ molecule
with the aug-cc-pVDZ basis set, and estimate the frozen-core FCI energy by
extrapolating the variational energy with respect to the renormalized
second-order perturbative correction.\cite{garniron_2019}
The curve is first computed using a simple determinant selection,
minimizing the energy without considering the spin operator. Then, 
the curve is computed using the spin-adapted determinant
selection, and we report in Fig.~\ref{fig:n2} the difference in
extrapolated FCI energies, as well as the overlap between the two wave
functions at each point of the curve.
For each point of the curve, the calculation was stopped when the wave
function was expanded on more than a million determinants. This
corresponds to second-order perturbative energy corrections smaller
than 0.012 hartree.

When the triple bond is broken, N$_2$ dissociates into two nitrogen
atoms, each in its high-spin configuration. At dissociation, the two
nitrogen atoms can be combined in a singlet, in a triplet or in a quintet
state, all with the same energy. Hence, without any particular
treatment all these spin states mix together and produce spin
contaminated wave functions. As the determinant-based Epstein-Nesbet
perturbation theory is not invariant with respect to the magnetic quantum
number $m_s$, spin contamination in the reference wave function
affects the second-order perturbative correction and makes the
extrapolations less accurate. This effect can be observed in
Fig.~\ref{fig:n2} where for distances larger than 2~{\AA} the overlap
between the spin-adapted singlet wave function and the determinant wave
function shows a significant spin contamination, leading to
fluctuations in the extrapolated energies as large as a millihartree.

\begin{figure}[ht]
  \begin{center}
  \includegraphics[width=0.48\textwidth]{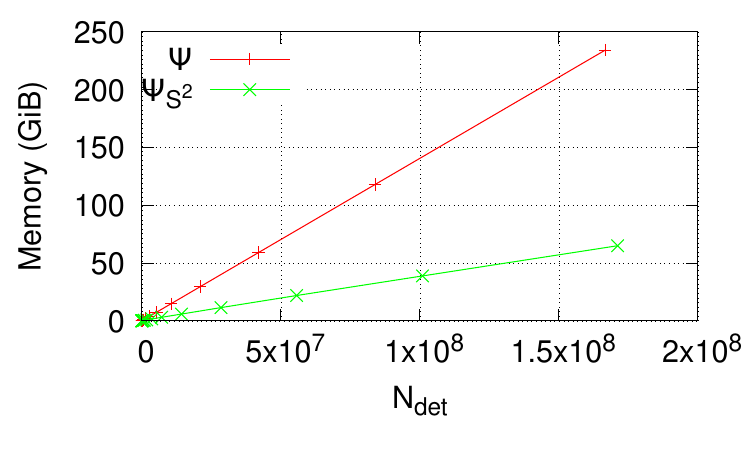}
  \includegraphics[width=0.48\textwidth]{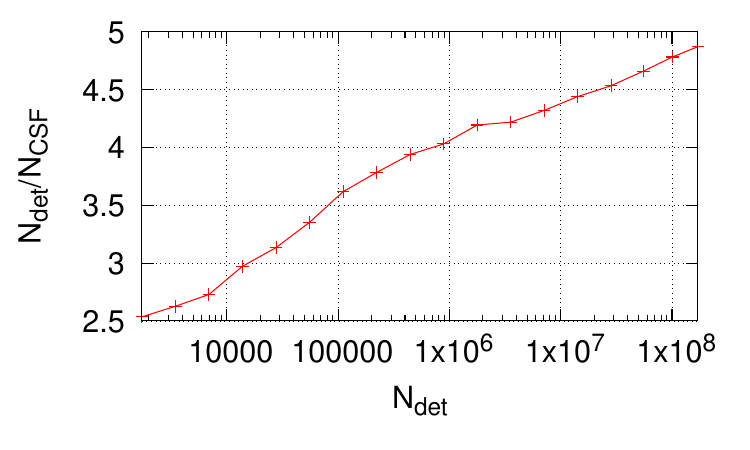}
  \end{center}
  \caption{Dissociated N$_2$ with the aug-cc-pVDZ basis.
    Top: memory requirements (in GiB) for the Davidson algorithm with storage
    in the determinant basis or in the CSF basis as a function of
    $N_{\text{det}}$.
    Bottom: Ratio of the number of determinants $N_{\text{det}}$ over the number of
    CSFs $N_{\text{CSF}}$, as a function of the number of selected determinants.}
  \label{fig:memory}
\end{figure}

Figure~\ref{fig:memory} shows the memory requirements of the Davidson
routines in the determinant-based and the CSF-based
storage. From this figure, it is clear that storing the matrices
in the CSF basis makes a big difference in terms of memory
requirements, with a reduction by a factor $4$ in the case of 
dissociated N$_2$ at an internuclear distance of 5~\AA with the
aug-cc-pVDZ basis.
This figure also displays the ratio $N_{\text{det}} / N_{\text{CSF}}$
as a function of $N_{\text{det}}$, which clearly tells us that the
number of determinants increases faster than the number of CSFs. This
can be explained by the fact that upon starting with a closed shell
reference, during the CIPSI selection, determinants with
a large number of open shells appear later than determinants with
mostly closed shells. Hence, we expect the reduction in terms
of memory requirements to be increasingly notable as the number of CIPSI
iterations increases.

\section{Conclusion}

We have presented a general algorithm to complement an arbitrary wave
function with all the required Slater determinants to obtain
eigenstates of the $\stwo$ operator when the Hamiltonian is
diagonalized, with negligible computational overhead. This spin
adaptation step is introduced after the determinant selection of the
SCI algorithm, and enables the possibility to switch to the
CSF basis for the diagonalization of the Hamiltonian to reduce the
memory requirements which is one of the limiting step is today's SCI algorithms.
We would like to emphasize that this spin-adaptation procedure can be
applied to any SCI-type method: CIPSI, SHCI, FCIQMC, \textit{etc}.
We hope to report further algorithmic improvements in the near future following the same philosophy.

\begin{acknowledgments}
  The authors gratefully acknowledge Sean Eron Anderson for creating
  the \emph{Bit Twiddling Hacks}\cite{NextBit} web page.

  This work is supported by the European Centre of Excellence in Exascale
  Computing TREX - Targeting Real Chemical Accuracy at the Exascale. This project
  has received funding from the European Union's Horizon 2020 - Research and
  Innovation program - under grant agreement no. 95216.
  AS and PFL were also supported by the European Research Council (ERC) under the 
  European Union's Horizon 2020 research and innovation program (Grant agreement No.~863481).
  Funding from Projet International de Coopération Scientifique
  (PICS08310) is acknowledged.
  KG acknowledges support from grant number CHE1762337 from the
  U.S. National Science Foundation.
  This research used resources of the Argonne Leadership Computing
  Facility, which is a U.S.  Department of Energy Office of Science User
  Facility operated under contract DE-AC02-06CH11357.
  This work was performed using HPC resources from CALMIP (Toulouse)
  under allocation 2021-18005 and from GENCI-TGCC (Grant
  2020-A0040801738).

\end{acknowledgments}


\bibliography{s2}

\end{document}